\documentclass{article}
\usepackage{arxiv}
\usepackage{amsmath,amsfonts}
\usepackage{algorithmic}
\usepackage{algorithm}
\usepackage{array}
\usepackage{textcomp}
\usepackage{stfloats}
\usepackage{url}
\usepackage{verbatim}
\usepackage{graphicx}
\usepackage{cite}

\usepackage{abstract} 
\usepackage{authblk}        
\usepackage{hyperref}
\usepackage[font=scriptsize]{subfig}
\usepackage{booktabs}       
\usepackage{placeins}       
\usepackage{threeparttable} 
\usepackage{tabularx}
\usepackage{xcolor,colortbl}
\definecolor{green}{RGB}{186,219,212}
\definecolor{red}{RGB}{225,180,171}
\definecolor{grey}{RGB}{229,229,229}
\definecolor{lila}{RGB}{129,15,124}
\definecolor{blue}{RGB}{4,140,212}
\definecolor{darkblue}{RGB}{4,114,178}
\definecolor{teal}{RGB}{0,144,160}
\definecolor{blue_plot}{RGB}{43,107,227}
\definecolor{yellow_plot}{RGB}{230,180,0}
\definecolor{green_plot}{RGB}{43,168,74}
\definecolor{blue_text}{RGB}{44,99,233}
\definecolor{yellow_text}{RGB}{200,150,0}
\definecolor{green_text}{RGB}{31,135,56}

\newcommand{\eyesonlyc}{{\color{blue_text}Eyes-Only}}
\newcommand{\mirrorsonlyc}{{\color{yellow_text}Mirrors-Only}}
\newcommand{\mirroreyesc}{{\color{green_text}Mirror Eyes}}

\newcommand{\eyesonly}{{Eyes-Only}}

\newcommand{\mirrorsonly}{{Mirrors-Only}}
\newcommand{\mirroreyes}{{Mirror Eyes}}
\newcommand{\mirroreye}{{Mirror Eye}} 
\newcommand{\mycolorbox}[2]{{\setlength{\fboxsep}{1pt}\colorbox{#1}{#2}}}

\title{Virtual Reflections on a Dynamic 2D Eye Model Improve Spatial Reference Identification}

\author[1, 2]{\href{https://orcid.org/0000-0002-2730-1116}{Matti Kr\"{u}ger}}
\author[1]{\href{https://orcid.org/0009-0002-1597-0547}{Yutaka Oshima}}
\author[1]{\href{https://orcid.org/0000-0003-1289-2504}{Yu Fang}}

\affil[1]{Honda Research Institute Japan Co., Ltd., Wako-shi, Saitama, Japan}
\affil[2]{Honda Research Institute Europe GmbH, Offenbach am Main, Germany}
\affil[ ]{\texttt{firstname.lastname@jp.honda-ri.com|@honda-ri.de}}
\date{}

\renewcommand{\shorttitle}{Virtual Reflections on a Dynamic 2D Eye Model Improve Spatial Reference Identification}

\hypersetup{
pdftitle={Virtual Reflections on a Dynamic 2D Eye Model Improve Spatial Reference Identification},
pdfauthor={Matti Kr\"{u}ger, Yutaka Oshima, Yu Fang},
pdfkeywords={eye-model, spatial referencing, human-robot interaction, virtual reflections, robot mediation, group interaction, robot-mediated interaction, word chain, shiritori, Georeferenced, Eye Model, Mental State, Eye Movements, Accurate Identification, User Experience, Spatial Attention, Human-machine Interaction, Interaction Scenarios, Attention Of People, Personality, Reaction Time, Visual Processing, Post-hoc Comparisons, Eye Contact, Positive Identification, Camera Images, Face Processing, Button Press, Environmental Elements, Mirroring, Interpupillary Distance, False Responses, Joint Attention, Gaze Direction, Objects In The Scene, Strong Attraction, Face Detection, Block Type, Post-hoc Multiple Comparisons
},
}

\begin{document}

\twocolumn[
\scriptsize{\textit{This article has been accepted for publication in IEEE Transactions on Human-Machine Systems. This is the author’s version of the work. It has not been fully edited and may differ from the final published version. Citation information: \href{https://doi.org/10.1109/THMS.2026.3651818}{DOI 10.1109/THMS.2026.3651818}}}
\maketitle
\begin{abstract}
The visible orientation of human eyes creates some transparency about people's spatial attention and other mental states. This leads to a dual role of the eyes as a means of sensing and communication. Accordingly, artificial eye models are being explored as communication media in human-machine interaction scenarios.
One challenge in the use of eye models for communication consists of resolving spatial reference ambiguities, especially for screen-based models. To address this challenge, we introduce an approach that incorporates reflection-like features that are contingent on the movements of artificial eyes. 
We conducted a user study with 30 participants in which participants had to use spatial references provided by dynamic eye models to advance in a fast-paced group interaction task. Compared to a non-reflective eye model and a pure reflection mode, the superimposition of screen-based eyes with gaze-contingent virtual reflections resulted in a higher identification accuracy and user experience, suggesting a synergistic benefit. \bigskip
\end{abstract}
]

\section{Introduction}
First and foremost, the eyes are sensory organs that acquire visual information from the environment. Sharp visual perception through the eyes relies on the projection of light from a scene at a depth of interest onto light-sensitive tissue inside the eye. To achieve such a projection selectively, the eyes have evolved mechanisms that allow us to sample a scene through various actions that alter, for instance, eye orientation, pupil diameter, and lens curvature. 
One byproduct of this action-guided visual perception is that aspects of cognition become observable. For example, eye movements and the direction of the pupil reveal where a person is looking and, when further following the gaze direction, what a person is currently attending to.  
Thus, eyes may be understood not only as sensory organs but also as communication devices \cite{kobayashi1997unique, Fang2024}.
Indeed, in support of their informative value, eyes have been found to be a particularly strong attractor for other people’s gaze (see, e.g., \cite{Langton1999,Langton2000,Itier2007}) and the exploitation of gaze direction transparency appears to even have evolved into reflexive gaze following behavior~\cite{Driver1999}. According to the cooperative eye hypothesis~\cite{kobayashi1997unique, Tomasello2007}, human eyes may have developed their salient features precisely to increase their social utility. 
The social role of the eyes is also emphasized by the participation of the eyes in the expression of and influence on emotions~\cite{Baron2001,Adolphs2005,Hietanen2008} as well as the particular sensitivity to eye contact \cite{Senju2009}.
Due to the perceptual salience of the eyes, communication abilities, and social role, models that mimic human eyes are regularly used as communication tools, especially in virtual characters and robots designed for human-machine interaction (e.g., \cite{Metta2008,Breazeal2008,Zaraki2014gazecontrol,admoni2017social,Gomez2018,Yoshida2022,Fang2024}). If implemented successfully, robotic eye models have been found to elicit similar psychophysiological responses in humans as the eyes of human interaction partners\cite{Kiilavuori2021roboteyecontact,Linnunsalo2023Intentionality}. Such findings suggest that eye models may be purposefully used in interactive and especially in cooperative~\cite{Krueger2017,Sendhoff2020,Bhaskara2020transparency} settings, e.g., to capture a person’s attention intuitively, communicate directions of interest, or indicate specific mental states.

\subsection{Challenges for communicative eye models}

However, eye models on virtual characters and robots are often subject to constraints that may limit communication utility. Recreating human-like ocular behavior requires multiple actuators to be driven in a complex conjunction at low latency and with high accuracy (e.g., ~\cite{teyssier2021eyecam}). This means that design and construction are either complex or a compromise in terms of model performance. 
Artificial eyes can therefore easily fail to recreate realistic eye movements with human-like latency, accuracy, and context-dependent variability, due to limitations of the available hardware or the behavior-driving models\cite{admoni2017social}. In consequence, their movements can appear unnatural, scripted, or even disturbing~\cite{Mori2012}, possibly leading to an ignorance of the eyes in the long term. 
Virtual eyes that are presented on displays (e.g.,~\cite{Gomez2018,Yoshida2022,fang2023designing,Fang2024Roman,admoni2017social}) may overcome many limitations related to latency and available degrees of freedom but can have difficulties in unambiguously conveying spatial information due to their two-dimensional screen projection. 

The ability to follow the gaze of another person is partially dependent on the spherical shape of the eyes, which produces visual cues that help estimate the orientation of each eye.
Considering the relative orientations of both eyes can reveal not just the direction but the actual distance and even the point of attention in space because the point of convergence of both eyes’ orientation (in extension) corresponds to the focus point~\cite{Howard1995}. Because the angle at the point of convergence becomes smaller with distance and, conversely, increases with proximity, this eye vergence angle (EVA) is a useful indicator of focus distance. The EVA is closely tied to the interpupillary distance (IPD), which varies slightly as a function of the EVA and thus encodes some depth information. In natural 3D settings, this information is reinforced by motion parallax, as small head movements produce dynamic changes in the observer's view that further disambiguate the target of another person's gaze.  

Estimating another person's eye orientation and focus direction becomes more challenging when the eyes are distant or presented in a 2D format, such as on screens. In these cases, contextual cues, such as head orientation and body posture, can support inferences about another person's attention region~\cite{Langton2000}. However, the lack of depth information can lead to misperceptions of gaze direction and distance, particularly in ambiguous situations like group interactions. While depth may be simulated on 2D displays through artificial motion parallax, e.g., via head-tracking contingent image changes~\cite{Lee2008, bates2018head}, such approaches are limited as they rely on observer movement and do not engage stereopsis. Moreover, adapting depth cues to individual perspectives becomes problematic in multi-person interactions, limiting its utility to one-to-one interaction scenarios.  

To overcome the loss of depth information while preserving the benefits of virtual eyes and 3D facial features, researchers have developed retro-projected robot heads~\cite{Delaunay2009}. These project images onto a translucent head model, creating a 3D display that conveys spatial information. However, this approach is susceptible to ambient illumination~\cite{Delaunay2009}, which can cause projected elements to be obscured and create unnatural lighting effects, such as inverted shadows. Retro-projected head and eye models may hence be restricted to environments with sufficiently controlled illumination. 

All discussed approaches for creating eye models for communication purposes come with issues that constrain their utility in various ways. The following section aims to introduce an approach that may circumvent some of these issues and extend the social utility of artificial eye models. 

\subsection{One-step spatial references} 
\label{sec_me_introduction}

In human interactions, understanding another person's focus of attention involves two steps: looking at their eyes and mapping their gaze onto its target in space. Uncertainty in ocular cues can make this spatial inference unreliable. However, artificial or virtual eyes are not bound by biological constraints but can be visually modified to reduce communicative uncertainty. This raises the question of how their appearance might be adapted to intuitively convey spatial references without interfering with other communicative functions. 

\begin{figure}[t]
    \centering
    \includegraphics[width=0.75\linewidth]{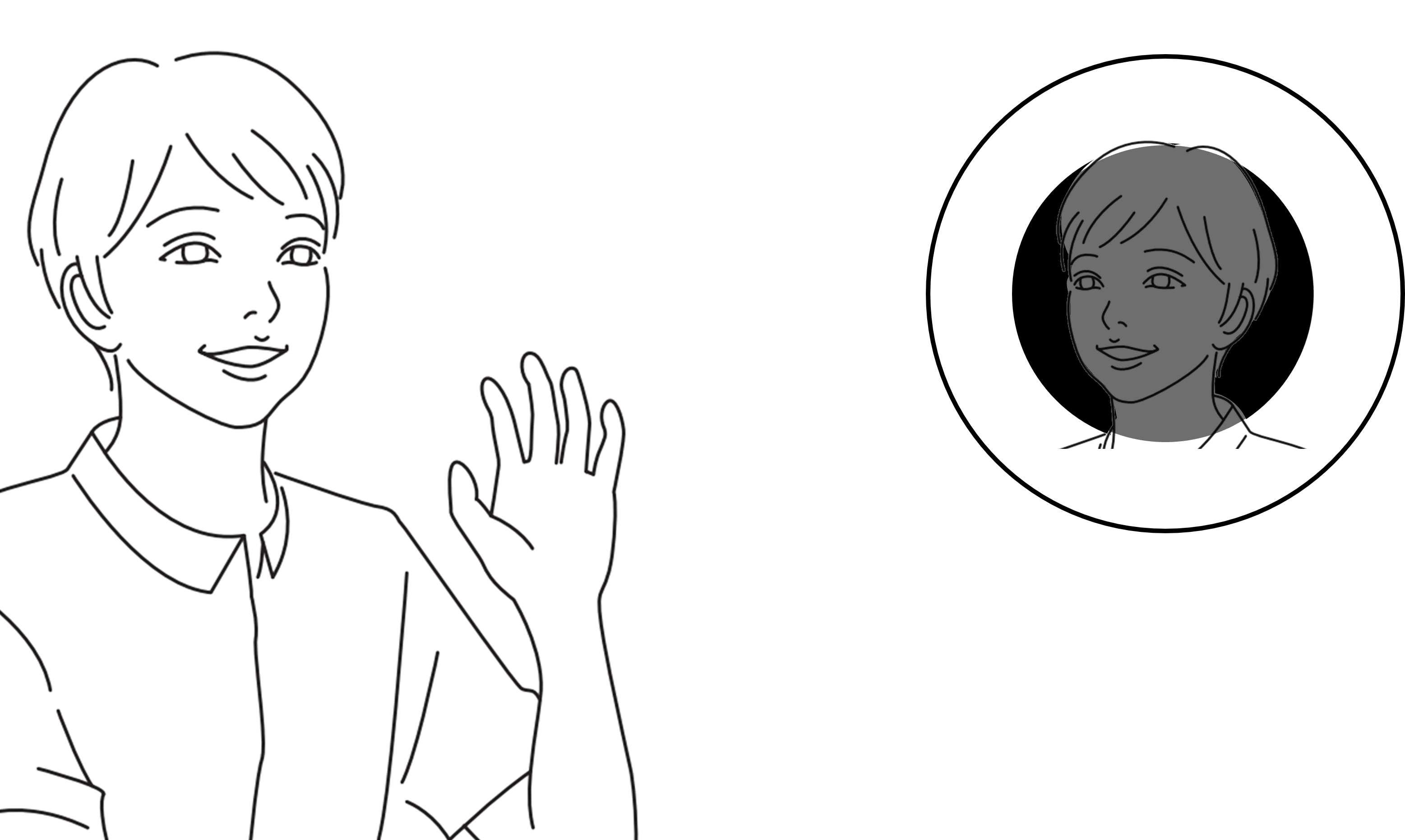}
    \caption{Illustration of an eye model that displays a mirror image of an attended area as a reflection-like overlay on its pupil and iris.}
    \label{fig:me_basic_illustration}
\end{figure}

Human eyes partially reflect their surroundings like convex mirror, although the small size of the reflection makes it difficult to recognize. In contrast, artificial eyes are not limited by size or physical reflection properties. 
Reflections on such eyes may be more recognizable, especially because the use of rendered eye models also allows for virtual reflections and projections that can be altered to enhance visibility. Consequently, here we propose to augment artificial eye models intended for human-machine interaction, such as eyes shown on screens, by including virtual mirror images of the machine’s current focus of attention in the eye models (see Figure~\ref{fig:me_basic_illustration}).  
In reference to that feature, we refer to such eye models as \textit{\mirroreyes}. 

\mirroreyes~simultaneously display the direction and content of attention. For an observer, this can turn the described two-step process of understanding spatial references into a single step and remove the need for unambiguously interpretable gaze vectors, substantially facilitating the design of effectively communicating eyes. Although the added detail displayed on such \mirroreyes~could increase information processing demands for an observer, we expect the perception of virtual reflections to be accompanied by a gain in gaze target identification accuracy, especially in situations with potential reference ambiguity, such as group interactions. 
We further expect the experience of such references to be enhanced in comparison to classical eye models and to pure dynamic virtual reflections that lack eye features, such as pupil and iris, which typically indicate spatial attention. 

To formalize the preceding claims, we propose the following research hypotheses. 
The use of \mirroreyes~...
\begin{description}
    \item[H1:] ...~ ~improves people's identification accuracy of spatial references compared to eye models that lack mirror augmentation.
    \item[H2:] ... reduces the speed of positive spatial reference identification compared to eye models that lack mirror augmentation. 
    \item[H3:] ...~ as a spatial reference tool in a group setting improves the user experience compared to eye models that lack either mirror-like augmentation or pupil-like features. 
\end{description}

To evaluate these hypotheses, we developed a functional prototype of an eye model with \mirroreye~capabilities and conducted a user study using this model. The remainder of this paper presents both the model and the empirical investigation in detail. 

\section{Methodology - Prototype}

\begin{figure}[t]
    \centering
    \includegraphics[width=1\linewidth]{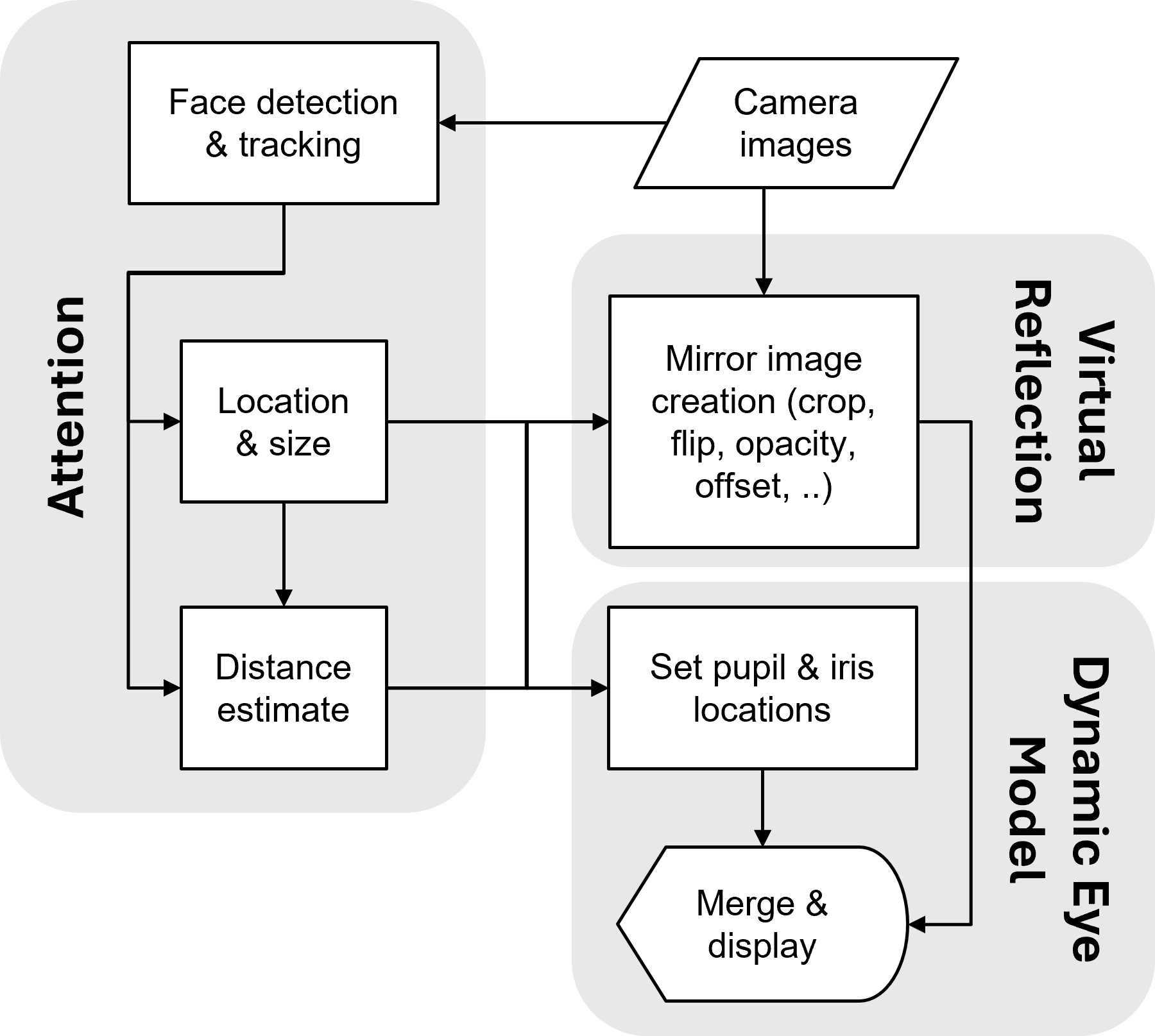}
    \caption{Connected modules of the \mirroreye~prototype.}
    \label{fig:submodules}
\end{figure}

We created a functional prototype that implements the \mirroreye~concept for utilization on a wide range of hardware, such as personal computers with webcams. It relies on the following interconnected modules as depicted in Figure~\ref{fig:submodules}:

\subsection{Dynamic eye model}
\label{sec_eye_model}
The first module consists of a dynamic 2D eye model that allows real-time movement of pupil and iris images relative to the sclera to create an illusion of rotation. To align these movements with locations of interest in space such that the pupil appears to point towards a target location, a mapping from a camera image space to eye coordinates is applied (see Figure \ref{fig:spatial_shift} and Equations~\ref{eq_ey}-\ref{eq_ex}). The camera image is mapped from the perspective of the location of the eye model on a screen. To simulate sensitivity to depth despite the flat camera image, we have created a module that estimates distances to objects in the scene based on provided object sizes and camera specifications for focal length and sensor size. The depth estimates are then used to adjust horizontal offsets in the case of the simultaneous application of multiple eye models. 
In accordance with typical vertebrate anatomy, we apply two eye models in conjunction and utilize the depth estimates to create individual horizontal offsets for each eye’s pupil and iris such that interpupillary distance varies with focus distance, resulting in an effect of binocular disparity. This allows an observer to infer depth information from alignment differences between the eyes and thus may facilitate gaze target recognition. 

\subsection{Attention}
\label{sec_attention_module}
To make the eye modules come to life by moving across a scene in a meaningful way, an attention module is required. Simply put, the task of such a module is to define and continuously update target coordinates in camera space that are then mapped to pupil and iris position as described above. 
Because we were interested in the application of eye models as referencing tools in social settings, we chose to let the eyes look at and follow people in a scene. For this purpose, we implemented a face-tracking module that can detect and simultaneously track multiple faces present in a camera feed. For the purpose of the present work, we based the face-tracking on the spatiotemporal proximity between individual face detections. With this method, faces detected at subsequent time steps near previous detections in image space are assigned the same identifier. This enables smooth tracking without the need for individual face learning or recognition for scenarios in which the minimal distance between multiple people can be well controlled. 
Target coordinates are then defined relative to one or multiple recent coordinates of face identifiers, allowing the eyes to look at, avoid, follow, and switch between faces. 

\subsection{Virtual reflection}
\label{sec_reflection_module}
\begin{figure}[t]
    \centering
    \includegraphics[width=0.8\linewidth]{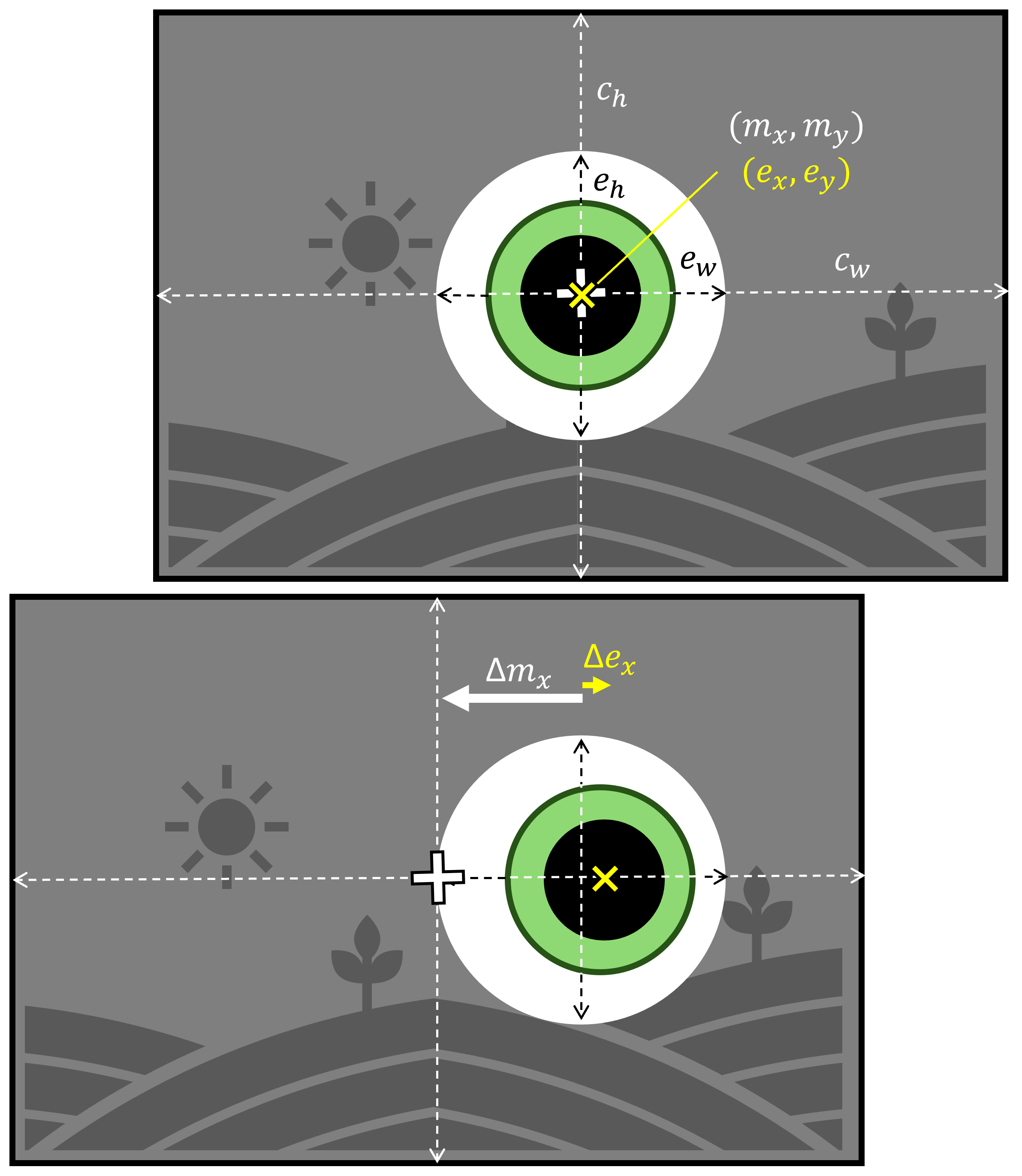}
    \caption{Illustration of the spatial shift for pupil and mirror image location as described by Equations \ref{eq_my}-\ref{eq_ex}. Top: All centers are aligned. Bottom: A small rightward shift of the pupil is accompanied by a larger leftward shift of the mirror image.}
    \label{fig:spatial_shift}
\end{figure}

Having established dynamic eye models that appear to look at and follow faces in a scene, the next module adds the virtual reflection capability consisting of the following steps: For each eye, an overlay of a horizontally flipped camera feed is created at a chosen transparency level. To indicate what part of the scene is being attended, the pupil should be located at or near the region of interest on the respective camera image. However, only moving the full eye model across the large image space to point to objects in the scene would result in a loss of the previously described eye-scene alignment. To overcome this issue, we apply a spatial shift and coordinate translation that moves the modified camera image opposite to the direction of eye movement as depicted in Figure \ref{fig:spatial_shift} and described by Equations \ref{eq_my}-\ref{eq_ex} as follows.

Given a camera image with size $c_w$ x $c_h$, an eye model with size $e_w$ x $e_h$, the spatial shift of the mirror image $(m_x, m_y)$ for each eye with spatial offset $(o_x, o_y)$ is given by 

\begin{equation}
\label{eq_my}
m_{y} = (\frac{c_y}{c_h} + o_y)\cdot(c_h-e_h) + \frac{e_h}{2}
\end{equation}
and 
\begin{equation}
\label{eq_mx}
m_{x} = (1-(\frac{c_x}{c_w} + o_x))\cdot(c_w-e_w)+\frac{e_w}{2}
\end{equation}

The location of the pupil and iris on the eye $(e_x, e_y)$ moving in the opposite direction is then given by  
\begin{equation}
\label{eq_ey}
e_{y} = c_{y} - m_{y} + \frac{e_{h}}{2}
\end{equation}

\begin{equation}
\label{eq_ex}
e_{x} = c_{x} - n_{x}(c_w-e_w)
\end{equation}

This dynamic reflection technique has the added benefit of mapping potentially large camera images to relatively small eye models without loss of resolution through the coordinate translations while contributing to an appearance of the eyes as three-dimensional reflectors for which the rotation magnitude is defined by the coordinate translation magnitude. 

\subsection{Software}
We implemented all modules using the Python programming language~\cite{python} with the libraries OpenCV~\cite{opencv_library} for image processing and MediaPipe~\cite{mediapipe} for face detection. 

\section{Methodology - Experiment}

To test the hypotheses specified in Section ~\ref{sec_me_introduction}, we designed an experiment that required participants within a group to quickly and accurately identify who in the group was being attended by an eye model to proceed in a fast-paced game-like task. In each experiment, we recorded the accuracy and speed of gaze reference identification. The purpose of the secondary game-like task was to create conditions that would allow a quick and structured recording of many samples while distracting the participants cognitively from the primary stimulus-response task. 

\subsection{Participant task}
\label{sec_participant_task}

As a secondary game-like task, we modified a traditional Japanese word game known as \textit{Shiritori}. 
The original game consists of taking turns saying words that start with the last letter of the word uttered by the previous person. For example: If person 1 says “Robot” then person 2 will need to say a word that starts with the letter “T” such as “trouble”, then it is again person 1’s turn to come up with a word that starts with “E” such as “ensures” followed by person 2 with a word starting with “S” and so on. 

For this study, we altered the rules of the game in the following way: 
\begin{description}
\item[1. Number of players:] The game is played with three people at a time.
\item[2. Unknown order:] The speaking order is unknown to participants for every word.
\item[3. Machine selection:] A machine mediator pseudo-randomly decides whose turn it is for each new word and then indicates its choice by looking at that person.
\item[4. Response button:] Participants who think that they have been selected need to press a button before being allowed to say the next word. They should press the button as soon as they feel that they are being selected and should think about the next word only after the button press. 
\item[5. Time limit:] There is a time limit of 3 seconds for pressing the button. After pressing the button there is a second time limit of 5 seconds for saying the next word.  If a person does not respond in time, the mediator will look at the next person. 
\item[6. Target succession rule:] The person who uttered the previous word cannot be immediately re-selected by the machine mediator unless the selected person fails to respond within one of the time limits.
\end{description}

The introduction of a third player and the machine mediator, instead of a turn-taking rule, ensured that participants were always uncertain about whose turn it would be next and had to make a decision based on the mediator's gaze to proceed with the game. 
The use of the button press was two-fold: It facilitated identifying and recording whether a participant had understood the gaze reference correctly and allowed us to measure how quickly true positive and false positive reference identifications took place. We considered the absence of a button press within the 3-second time limit as a negative response. 
The short time limit for the button press urged participants to react as quickly as possible, thereby reducing the potential influence of confounding factors and enabling rapid data collection.
We defined each sequence of gazing, button press, and word utterance as one trial. Thus, a single trial took a maximum of 8 seconds. 

\subsection{Experiment structure}
\label{sec:experiment_conditions}

\begin{figure}[ht]
\centering
    \subfloat[\eyesonlyc\label{fig:NoMirrorw}]{
    \centering
        \includegraphics[width=0.29\linewidth]{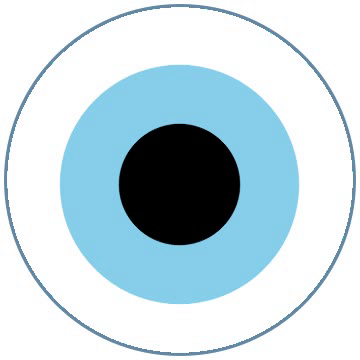}
    }
    \hfill
    \subfloat[\mirrorsonlyc\label{fig:MirrorOnlyBlurw}]{
    \centering
        \includegraphics[width=0.29\linewidth]{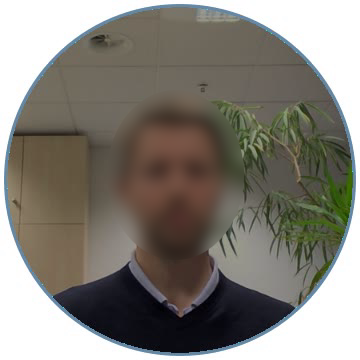}
    }
    \hfill
    \subfloat[\mirroreyesc\label{fig:mirroreyew}]{
    \centering
        \includegraphics[width=0.29\linewidth]{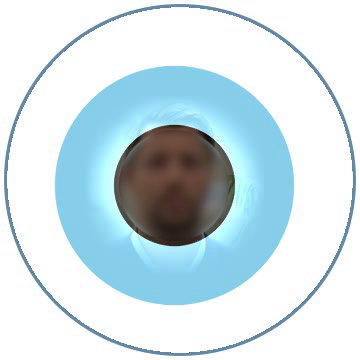}
    }
    \caption{Eye model display conditions. \ref{fig:NoMirrorw}: Basic eye model with no reflection of an attended face. \ref{fig:MirrorOnlyBlurw}: dynamic mirror that tracks an attended face. \ref{fig:mirroreyew}: Combination of eye model and dynamic mirror of attended face on pupil and iris (\mirroreyesc). The faces in Figures \ref{fig:MirrorOnlyBlurw} and \ref{fig:mirroreyew} have been blurred for publication purposes to protect privacy. No face blur was present in the experiment.}    
    \label{fig:expconditions}
\end{figure}

We utilized three different display conditions of the eye models as depicted in Figure~\ref{fig:expconditions}. The first condition (\eyesonlyc) consisted of pure eye models as described in Section \ref{sec_eye_model} without any reflection-like properties. The second condition (\mirrorsonlyc) consisted of a pure dynamic mirroring mode that was centered on the current face of interest as described in Section \ref{sec_reflection_module} but contained no typical eye features such as pupil and iris. 
The third condition (\mirroreyesc) consisted of the combination of the first two conditions, i.e., the overlay of dynamically updated mirror-like features on the pupil and iris. 
The use of these three conditions allowed us to test whether any potential effects of the \mirroreyes~might be attributed to one of its two major constituents or whether these would arise only from their combination. 

\subsection{Procedure}

\begin{figure}[t]
    \centering
    \includegraphics[width=0.9\linewidth]{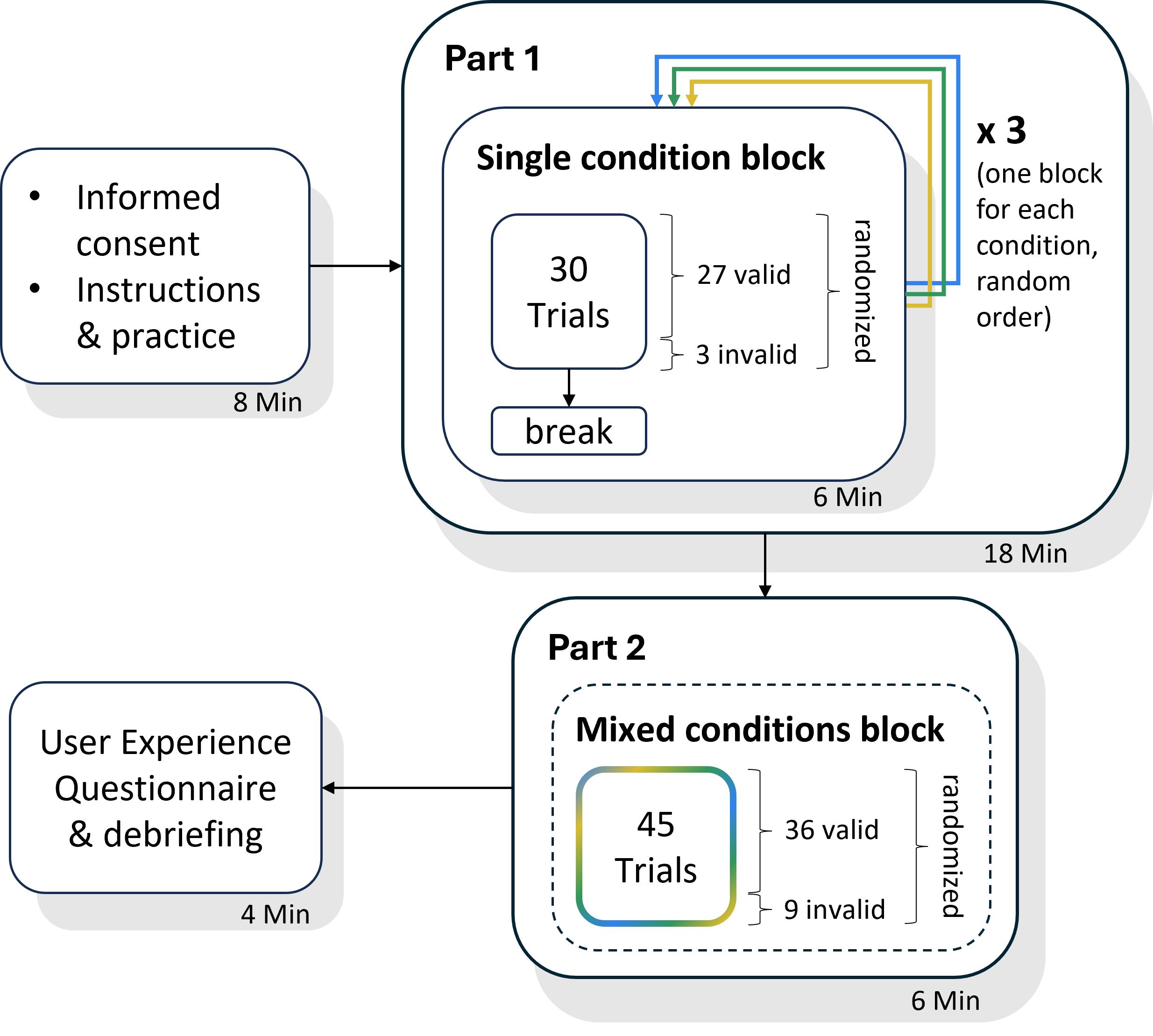}
    \caption{Experiment procedure with duration estimates for individual steps.}
    \label{fig:procedure}
\end{figure}

Figure \ref{fig:procedure} illustrates the procedure and individual components of the experiment.
Before starting the experiment, all participants received detailed instructions and practiced for 10 trials in each eye condition. 

The subsequent experiment consisted of two parts that differed in how trials of the different eye model display conditions were grouped into blocks. In the first part, each display condition was used for a block of 30 trials, resulting in three blocks. Blocks of this type are further referred to as "single condition block". 
The order of single condition blocks within the first part of the experiment was pseudo-randomized between participant triplets. 
The second part of the experiment consisted of a single "mixed condition block". In a mixed condition block the eye model display conditions switched pseudo-randomly between individual trials, such that the eyes might, for instance, look at one person with \mirroreyes~but focus on the next with non-reflective eyes. At least 15 trials of each condition were distributed across a mixed condition block. 
While the single condition blocks in the first part allowed participants to anticipate the nature of the cue they would receive from the mediator, the mixing of conditions in the second part added cue uncertainty, making tuning to a specific set of features disadvantageous. Differences between the two parts could thus point towards a possible dependence or independence between identification performance and cue type predictability. 

Because participants could \textit{steal} individual trials from each other by wrongly pressing the button when another participant was attended, we created a real-time trial balancing procedure that would either swap or append trials for involved participants in such cases depending on compatibility with the target succession rule (see Section \ref{sec_participant_task}). False responses by participants could therefore extend the total number of trials in a block up to a predefined total time limit of 4 minutes for the first part and 6 minutes for the second part. 

Assuming a medium effect size of $\eta^2 = 0.06$, a power analysis for a repeated measures design with three within-subject conditions, 30 participants, and an alpha level of 0.05 estimated the statistical power to be 84.9\%. 

\begin{figure}[t]
\centering
 \subfloat[Experiment stimuli presented on a TV display\label{fig:Exp_Stimuli}]{%
 \includegraphics[width=0.8\linewidth]{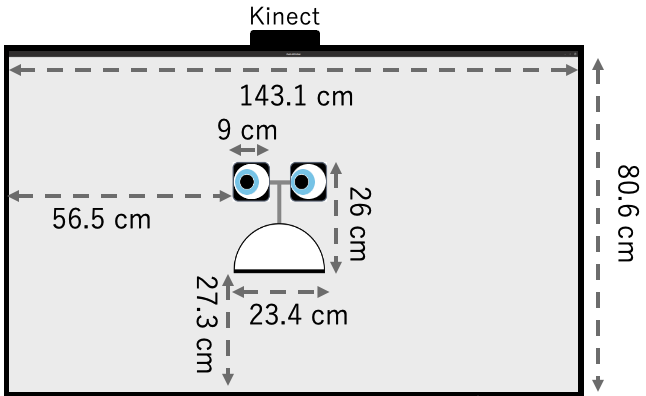}}\\
\subfloat[Schematic of participants' positions\label{fig:Exp_Setup}]{%
 \includegraphics[width=0.8\linewidth]{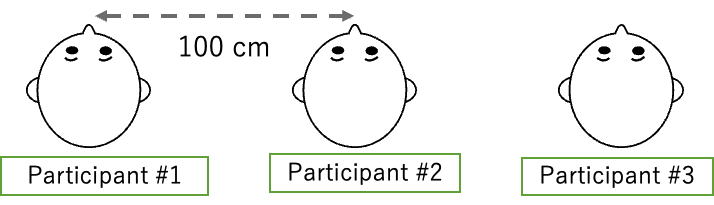}}\\
 \caption{Schematic of the experimental setup: \ref{fig:Exp_Stimuli}: Arrangement and dimensions of the visual stimuli consisting of one of the three dynamic eye models rendered into the eyes of a mock-up robot head shown on a gray display. \ref{fig:Exp_Setup}: Distancing between the three participants. The participants were free to move their heads and bodies while in their positions.}  
    \label{fig:expsetup}
\end{figure}

\subsection{Dependent measures}

\subsubsection{Reference identification performance} 

To test H1 and H2, we used the reference identification accuracy and reaction times as dependent measures. Both are derived from the participants' button press behavior. 
The reference identification accuracy for each participant is defined as binary classification accuracy, i.e., as the ratio of the sum of true positive ($TP$) and true negative ($TN$) identifications over the sum of all trials involving that participant. 

\begin{equation}
\label{eq_accuracy}
    \text{Accuracy} = \frac{TP+TN}{TP+TN+FP+FN}
\end{equation}

Therefore, the accuracy does not only take the true positive identification of a reference into account but is also affected by the participants' ability to identify when they are not being attended. To provide opportunities for the identification of such non-references, we programmed the eye model to "make mistakes" in a subset of trials by not attending to any particular participant but instead focusing on areas located between participants. For such trials, the lack of a button press within the time limit would then count as a true negative identification. To ensure exposure of each participant to at least one error case for each block and condition while minimizing error predictability, the rate of machine mistakes was set to 10\% (3 trials) for each condition block in experiment part 1 and to 20\% (3x3 trials) for part 2, which contained all three conditions. 

We chose to use only a single response button per participant to avoid any coordination-dependent confounders. This button indicated a positive response. Negative responses were given implicitly by not responding within the 3-second time limit. Therefore, reaction times were only available for true positive and false positive responses as indicated by Hypothesis 2.  

\subsection{User experience} 
In this experiment, the 8-item version of the User Experience Questionnaire (UEQ-S; \cite{schrepp2017design}) was used to assess participant experience with the virtual eyes. After completing all experimental blocks, participants filled out the UEQ-S, which consists of eight items, four representing a pragmatic scale and the other four representing a hedonic scale. Each item is rated on a 7-point semantic differential scale, ranging from -3 (most negative evaluation) to +3 (most positive evaluation). Among psychometrically validated tools for measuring user experience (e.g.,~\cite{brooke1996sus, Finstad2010umux, schrepp2017design}), we selected the UEQ-S because it efficiently captures two core user experience dimensions, the pragmatic and hedonic quality~\cite{hassenzahl2010needs}, while minimizing participant burden through its brief completion time.

\subsection{Participants}

The experiment involved 30 participants (4 females and 26 males; aged 21 to 56 years; mean age: $32.8 \pm 8.52$ years) with normal or corrected-to-normal vision, who participated in groups of three. Participants had no prior exposure to the experimental setup or eye model. All participants signed consent forms, and the experiment was reviewed and approved by the Bioethics Committee in Honda's R\&D (100HM-064H). 
One participant was excluded from accuracy and reaction time analysis due to a lack of recorded button presses.

\subsection{Setup}

Figure \ref{fig:expsetup} shows the schematic of the experimental setup. The experiments were conducted in an office environment in which three participants stood in front of a 65-inch display (Bravia KJ-65X80J; Sony, Japan). Participants stood approximately 2 meters from the display, with a 1-meter distance between one another (Fig. \ref{fig:Exp_Setup}). The positions were selected to provide a comfortable social distance while still keeping participants close enough to each other and far enough from the display to create some gaze direction ambiguity.  
A camera (Azure Kinect DK; Microsoft, USA) was mounted directly above the center of the screen to capture the scene containing the participants, i.e., the eyes' attention targets. Each participant was equipped with a key button (FS1-P; Edikun, Japan) on their hand, connected to the experimental computer via USB. A computer (Magnus One ECM73070C; Zotac, Hong Kong) managed both the visual stimulus presentation and data collection. The experimental stimulus, an image of a generic robot head equipped with the dynamic eye model of the respective condition, was displayed at the center of the screen on a gray background with a height of 26 cm (Figure \ref{fig:Exp_Stimuli}). Each eye was represented in a 9 × 9 cm square. The diameter of the sclera of each eye was 8.8 cm, the diameter of the iris (outer circle) was 6.4 cm, and the diameter of the pupil (inner circle) was 3.5 cm. The reflection opacity in the \mirrorsonlyc~and \mirroreyesc~ conditions was set to 60\%. The display measured 80.6 cm in height and 143.1 cm in width and was positioned on a stand that was about 99.5 cm tall. This positioning ensured that the eye models were at a height of approximately 1.5 meters, which allowed participants to observe them comfortably from a standing position.

\subsection{Software}
We implemented the experiment using the Python programming language~\cite{python}. The subsequent data analysis was conducted in Python using the libraries Pandas ~\cite{pandas} for data structuring, Matplotlib~\cite{matplotlib} and seaborn~\cite{seaborn} for visualizations, and statsmodels~\cite{statsmodels}, scikit-posthocs~\cite{scikit-posthocs}, and pingouin~\cite{Vallat2018} for statistical analysis. 

\FloatBarrier
\section{Results}

\begin{table}[h]
\centering
  \caption{Summary of Two-Way ANOVA results for accuracy values.}
  \label{tab:anovaacc}
\resizebox{\columnwidth}{!}{%
\begin{tabular}{lrrrrr}
\toprule
 & df & sum sq & mean sq & \textit{F} & $PR(>F)$ \\
\midrule
Display condition & 2 & 0.5430 & 0.2715 & 14.9880 & \textbf{$<$~0.0001} \\
Block type & 1 & 0.0009 & 0.0009 & 0.0512 & 0.8212 \\
Interaction & 2 & 0.0032 & 0.0016 & 0.0893 & 0.9146 \\ 
Residual & 168 & 3.0434 & 0.0181 &   &   \\
\bottomrule
\end{tabular} 
}
\end{table}

\begin{figure}[h]
\vspace{-0.5cm}
\centering
\subfloat[Single condition blocks (part 1)\label{fig:accuracy_single}]{%
 \includegraphics[width=0.49\linewidth]{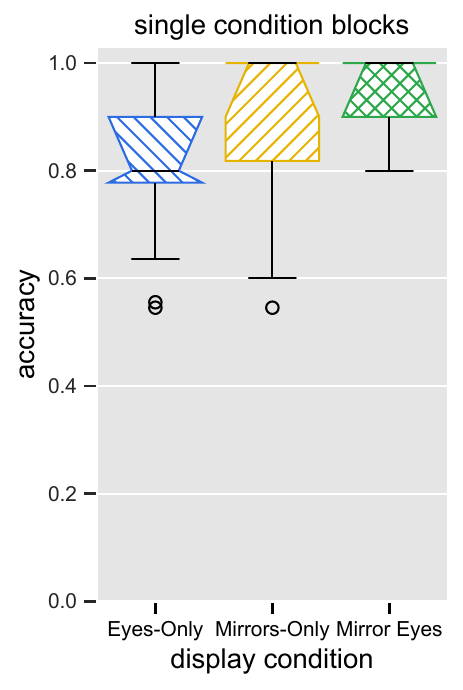}}
 \hfill
 \subfloat[Mixed condition block (part 2)\label{fig:accuracy_mixed}]{%
 \includegraphics[width=0.49\linewidth]{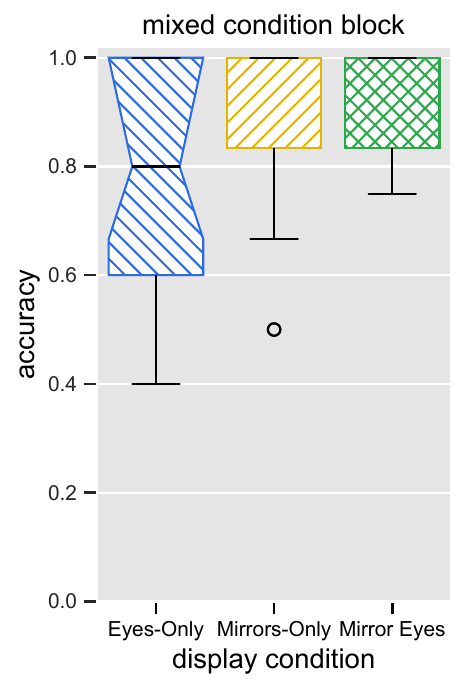}}
 \caption{Box plots showing the distributions of accuracy scores across conditions for single (left) and mixed condition blocks (right).}    
    \label{fig:accuracy}
\end{figure}

\begin{table}[t]
\centering
  \caption{Summary of Two-Way ANOVA results for reaction times. }
  \label{tab:anovart}
\resizebox{\columnwidth}{!}{%
\begin{tabular}{lrrrrr}
\toprule
 & df & sum sq & mean sq & \textit{F} & $PR(>F)$ \\
\midrule
Display condition & 2 & 0.0534 & 0.0267 & 0.3484 & 0.7063 \\
Block type & 1 & 0.1876 & 0.1876 & 2.4489 & 0.1195 \\
Interaction & 2 & 0.2440 & 0.1220 & 1.5921 & 0.2066 \\ 
Residual & 168 & 12.8724 & 0.0766 &   &   \\
\bottomrule
\end{tabular}
}
\end{table}

\begin{figure}[t]
\vspace{-0.5cm}
\centering
\subfloat[Single condition blocks (part 1)\label{fig:rt_single}]{%
 \includegraphics[width=0.49\linewidth]{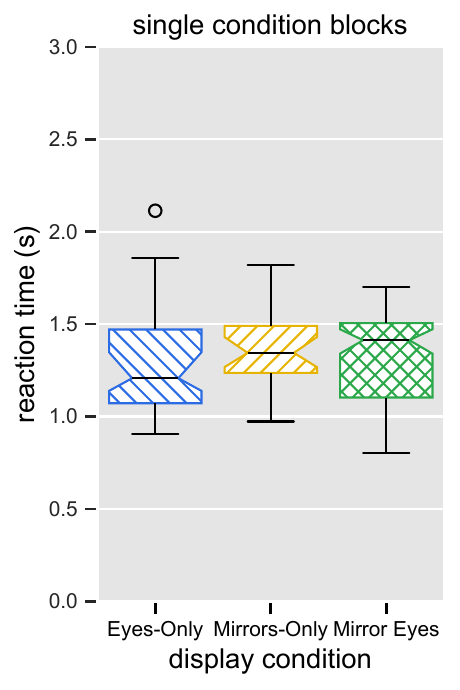}}
 \hfill
 \subfloat[Mixed condition block (part 2)\label{fig:rt_mixed}]{%
 \includegraphics[width=0.49\linewidth]{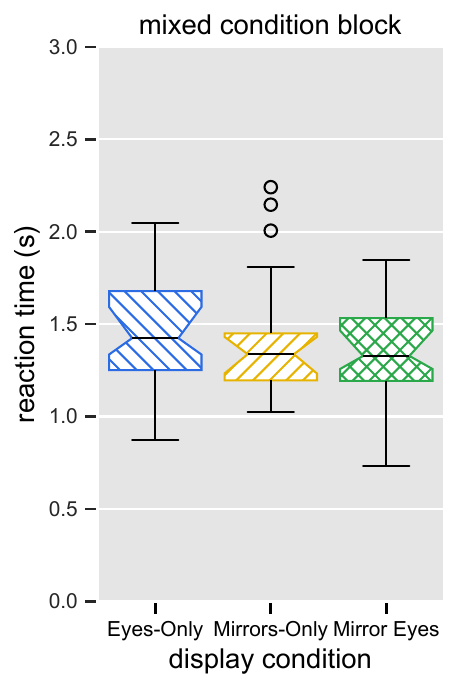}}
 \caption{Box plots showing the distributions of reaction times across conditions for single (left) blocks and the mixed condition block (right). }    
    \label{fig:rt}
\end{figure}

\subsection{Reference identification performance}

\subsubsection{Accuracy}

Figure~\ref{fig:accuracy} shows the distribution of accuracy values across conditions in both parts of the experiment. The accuracy was high across all conditions for the single condition blocks (\eyesonlyc: $\mu=0.82, sd=0.13$; \mirrorsonlyc: $\mu=0.91, sd=0.12$; \mirroreyesc: $\mu=0.94, sd=0.07$). 
Also in the mixed condition block the accuracy values were high across conditions (\eyesonlyc: $\mu=0.81, sd=0.20$; \mirrorsonlyc: $\mu=0.92, sd=0.15$; \mirroreyesc: $\mu=0.94, sd=0.09$).
A two-way ANOVA (see Table~\ref{tab:anovaacc}) revealed a significant main effect for display condition ($F(2,168)=14.99, p<0.001$) but no effect for block type ($F(1,168)=0.05, p=0.82$), or interaction between display condition and block type ($F(2,168)=0.09, p=0.91$). 
Post-hoc comparisons with Holm-Bonferroni adjusted p-values revealed a significant difference between the \eyesonlyc~and \mirrorsonlyc~($p < 0.01$) and between the \eyesonlyc~and the \mirroreyesc~condition ($p < 0.001$). 
These results support our first hypothesis and additionally indicate an accuracy advantage of a dynamic mirroring mode over pure eye-based referencing. There are no indications of an influence of cue type predictability on the average accuracy. 

\subsubsection{Reaction times}

Figure \ref{fig:rt} shows the distribution of reaction times across conditions in experiment parts 1 and 2. Average reaction times were similar across all conditions for the single condition blocks of experiment part 1 (\eyesonlyc: $\mu=1.29, sd=0.29$; \mirrorsonlyc: $\mu=1.38, sd=0.20$; \mirroreyesc: $\mu=1.34, sd=0.24$). 
Reaction times in the mixed condition block of experiment part 2 were comparable to those of the first part, especially for the two mirror conditions (\eyesonlyc: $\mu=1.46, sd=0.32$; \mirrorsonlyc: $\mu=1.40, sd=0.32$; \mirroreyesc: $\mu=1.35, sd=0.27$). 
A two-way ANOVA (see Table~\ref{tab:anovart}) did not reveal any significant effects for display condition ($F(2,168)=0.35, p=0.70$), block type ($F(1,168)=2.45, p=0.12$), or interaction between display condition and block type ($F(2,168)=1.59, p=0.20$). 

These results are not indicative of any reaction time disadvantage of the \mirrorsonlyc~and \mirroreyesc~modes compared to the \eyesonlyc~mode and hence do not support the second hypothesis. 
There are no clear indications of an influence of cue type predictability on the average reaction time.

\subsection{User experience} 

\begin{table}[t]
\centering
  \caption{Score means and standard deviations of the UEQ-S (N = 30)}
  \label{tab:UEQ_score}
  \vspace{-1.2pt}
  \resizebox{\columnwidth}{!}{%
  \begin{tabular}{lccc}
    \toprule
    Display condition & Pragmatic quality & Hedonic quality & Overall\\
    \midrule
    \eyesonlyc & 0.67 (1.27) & -0.26 (1.34) & 0.20 (1.12)\\ 
    
    \mirrorsonlyc & 0.71 (1.59) & -0.22 (1.46) & 0.25 (1.12)\\ 
    
    \mirroreyesc & 1.29 (1.38) & \textbf{1.30 (0.92)} & \textbf{1.30 (0.98)} \\
  \bottomrule
\end{tabular}
}
\vspace{1.8pt}
\scriptsize \newline
Bold values denote statistically significant differences ($p < 0.01$).

\end{table}

\begin{figure}[t]
    \centering
    \includegraphics[width=\linewidth]{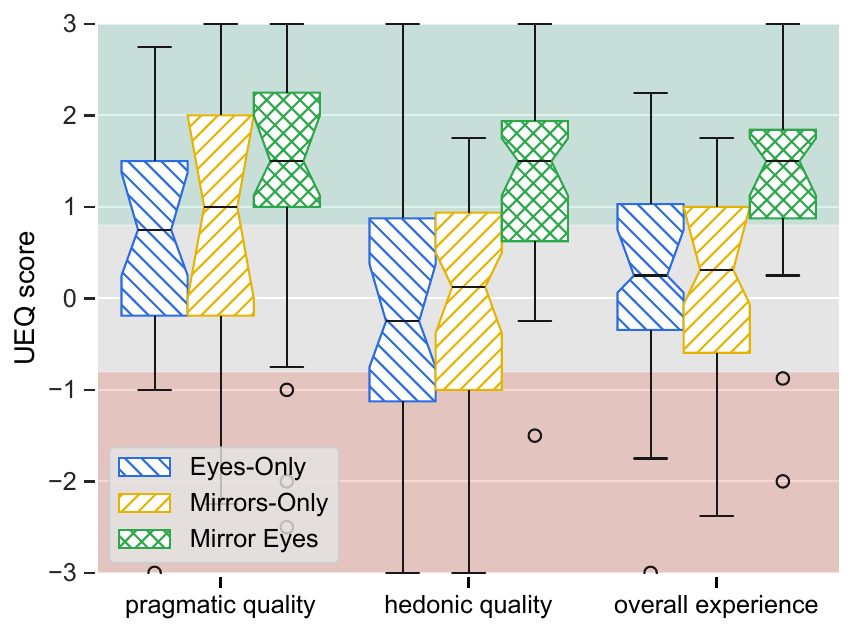}
    \caption{Box plots for UEQ-S rating of pragmatic and hedonic quality, and overall user experience across the three different conditions of the eye models. Different background colors represent evaluation ranges: values between $-0.8$ and $0.8$ indicate neutral ratings (\mycolorbox{grey}{grey}), values above $0.8$ signify positive ratings (\mycolorbox{green}{green}), and values below $-0.8$ indicate negative ratings (\mycolorbox{red}{red}).}
    \label{fig:UEQ}
\end{figure}

Figure \ref{fig:UEQ} presents the results of the UEQ-S ratings of 30 participants who evaluated the user experience across three different display conditions of the eye models, which were referred to as "Eyes", "Mirrors", and "Mirror Eyes" during the study. 
Table~\ref{tab:UEQ_score} shows the scale means for the UEQ-S ratings. Standard deviations are shown in parentheses. 

For the \eyesonlyc~condition, the mean score was 0.67 for pragmatic quality and -0.26 for hedonic quality. These scores are similar to those in the \mirrorsonlyc~condition for which the means were 0.71 for pragmatic quality and -0.22 for hedonic quality. In contrast, the \mirroreyesc~condition yielded the highest rating among the three conditions, with mean scores of 1.29 for pragmatic quality and 1.3 for hedonic quality. A one-way ANOVA revealed a significant difference in hedonic quality across the three conditions ($F(2,87) = 14.86, p < 0.001$). Post-hoc multiple comparisons with Holm-Bonferroni adjusted p-values showed a significant difference between the \mirroreyesc~and the other two conditions (both $p < 0.001$). For pragmatic quality, while the \mirroreyesc~condition had a higher mean (1.29, indicating a positive rating in the green zone), no significant differences were found between the three conditions ($F(2,87) = 1.81, p = 0.17$).

For the overall user experience evaluation, the \mirroreyesc~condition received a mean score of 1.30, while the \eyesonlyc~and \mirrorsonlyc~conditions scored 0.20 and 0.25, respectively. ANOVA results indicated a significant difference in overall user experience across the three conditions ($F(2,87) = 9.93, p < 0.001$). Post-hoc multiple comparisons with Holm-Bonferroni adjusted p-values confirmed that the \mirroreyesc~condition differed significantly from the other two conditions (both $p < 0.001$). The internal consistency of the pragmatic quality and hedonic quality scales for the \mirroreyesc~condition was sufficiently high, with Cronbach Alpha values of 0.88 (pragmatic quality) and 0.81 (hedonic quality). 

In sum, the results of the UEQ-S ratings indicate a more positive user experience for the \mirroreyesc~condition than for the other two conditions and thus support our third hypothesis.   
\section{Discussion}

Here, we introduced an approach for overcoming spatial referencing challenges of 2D eye models as communication tools in human-machine interaction. By augmenting a screen-based eye model with a dynamic reflection-like overlay that is contingent on eye movements, pupil-based pointing is complemented by direct scene context to improve spatial reference identification for observers. We conducted a user study with 30 participants who were asked to carry out a group interaction task that relied on an identification of spatial references produced by screen-based eye models to compare this "\mirroreyes" model to a pure eye model without mirroring capabilities and a pure dynamic mirror model without eye-like features. 

\subsection{Reference identification accuracy and speed}
Although accuracy was high for all tested eye models, the two eye models with mirror-like features led to significantly higher reference identification accuracy than a pure non-reflective eye model. Despite the added contextual information, the observers of augmented eye models needed to process, the study did not reveal any significant impact of mirroring on reaction times. The lack of noticeable impact on reaction times stands in contrast to our second hypothesis, which assumed a speed reduction for reference identification as an expression of a potential trade-off for the additional information that needs to be processed. The absence of a mirroring-related speed trade-off might in part be explained by the practice trials prior to the experiment. These could have given participants an opportunity to identify and focus on the most informative features. 
Future work should facilitate the separation of dependent measures from early and later trials of each experimental condition to identify potential learning effects. 

Another possible explanation for the lack of an apparent trade-off lies in the reference target class used in our study: The eye models always focused on faces. 
Human brains have evolved highly dedicated face processing abilities~\cite{Mccarthy1997}, likely due to their social role, allowing us to recognize target faces after as little as~260 ms~\cite{besson2017}. Furthermore, the seen gaze direction has been found to modulate the activity of regions associated with face processing~\cite{George2001}, which could suggest that a linkage between the two, as introduced here, might be exploiting fast visual processing networks. In addition, occasionally seeing their own faces might have facilitated the processing of other participants' faces~\cite{Li2011}, which could have reduced the risk of false positive identifications. 

However, the potential links to neural face processing and associated findings also highlight one limitation of the present study: Because faces were the only spatial reference targets, we gain no insights about the utility of the approach for conveying references to other object categories that are likely to have fewer dedicated neural processing resources. Future investigations should therefore expand the set of gaze target object categories.

\subsection{User experience}
Our findings regarding accuracy and response speed are consistent with user experience ratings obtained via the UEQ-S questionnaire.  
The UEQ-S results indicate that the \mirroreyes~condition was attributed significantly higher hedonic qualities than the other conditions, supplying the means for the step from a neutral to a positive user experience. Overall, the statistical analysis revealed that the user experience with the \mirroreyes~condition was significantly better than with the \eyesonly~or \mirrorsonly~conditions. 
The participants hence appear to have largely appreciated the support given by the \mirroreyes~condition. 
    
Notably, a single participant differed in this regard by rating both mirror conditions lower than the pure eye model. This participant explained afterwards, that seeing oneself in the eyes evoked a strong sense of being observed that made them feel uncomfortable. Although this response came from a 1/30 minority of the participants, the potential for such responses may require additional consideration in future interaction designs involving similar features. Furthermore, the male-skewed participant sample may have constrained the range of experiences and behaviors represented in our findings. 

\subsection{Mental state transparency}

Differential emotional responses to seeing oneself in the mirror have been described before~\cite{Mulkens2009,Barnier2019,Ypsilanti2020} and could account for some variability in user experience. 
Another factor that may contribute to an adverse sense of being observed is an attribution of judgment or malevolence to the observing entity. Because we were mainly interested in the \mirroreyes' utility for improving spatial ocular referencing, we kept the expression of the eye model neutral. However, we cannot rule out the possibility that individual users may attribute diverse mental states or personalities~\cite{Luo2019roboticeyespersonality} to the agent driving an eye model. 
Close associations between eyes and mental states may impair some interaction scenarios, but also highlight further potential in the utilization of augmented eye models such as the \mirroreyes.

Often described as windows to the mind, the eyes can reflect a larger set of mental states than just spatial attention. For instance, carefully designed gaze patterns, especially when integrated with contextual information in avatars or virtual agents, have been used in VR to subtly convey cognitive states and enhance social presence~\cite{tao2023embodying,Ruhland2015}. Moreover, when combined with non-verbal cues such as gestures and facial expressions, gaze inference models show great promise in human-computer interaction applications \cite{yamashita2024personality, gomez2021developing}, and can also be used to express personality traits~\cite{yamashita2024personality} or emotional states like fear and anxiety~\cite{oliveira2024differential}. 
Nevertheless, close associations between eyes and mental states may also limit possible applications. 

\subsection{Cue saliency and attention}

Eyes by themselves are strong attention attractors~\cite{Langton1999,Langton2000,Itier2007}. Reflection-like overlays may increase the salience of eye models even further, especially for reflections of other strong attractors like faces~\cite{Langton2008faces,Gliga2009faces}.
The salience of any machine expression deserves consideration in interaction design because it affects the balance between observability and distraction. Typically, a high expression salience makes an observer more likely to look at a face and less likely to look at elements in the environment. 
But because the virtual reflections are references to elements in the machine's environment, their effects on the distribution of an observer's attention could be twofold: 
They could draw attention to the eyes themselves, and they could draw attention to the reflected element in the environment. 
\mirroreyes~could therefore serve as a tool to establish joint attention between machine agents and human observers.
On top of their original role as a general purpose tool for spatial references, an ability to establish and modulate joint attention could extend means for machine-mediated joint attention skill development~(e.g., \cite{Zheng2013robotmediatedjointattention}) in people who experience challenges in reading social signals, as seen in individuals on the autism spectrum~\cite{dawson2004early,charman2003jointattention}.
Future work should therefore investigate the impact of the \mirroreyes~on people's attention and whether mirroring can be used for a selective modulation of eye model salience and gaze target salience. 

\subsection{References in physical space}

Looking back at the original objective of the present investigation, i.e., the improvement of spatial reference communication with 2D eye models, one aspect that requires closer examination is depth of gaze. Although our eye model encodes focus distance in the interpupillary distance and reflections naturally become smaller with increasing distance, the present study did not investigate the effects of Mirror Eyes at different levels of focus depth. 

Another direction for follow-up research can arise from an application of the approach within more physically instantiated machine interaction partners, such as robots that possess a movable head.  

Previous work has shown that humans perceive the direct gaze of a robot more clearly when the robot's eye gaze is coordinated with its head movement, as opposed to when the head remains stationary \cite{Fang2024,Fang2024Roman}. In the present study, all three eye conditions dynamically tracked the user's face, and the results demonstrate that participants could accurately judge the machine mediator's gaze direction, despite its static screen-based mock-up head. Previous research also suggests that combining eye and head movements while simulating visual perception processing enhances the perception of a robot's attentional gaze~\cite{Fang2024}. Also, potential cross-modal interactions with other output modalities such as speech (e.g., ~\cite{Srinivasan2014headgazespeech}) should be considered in future developments.

Moving forward, we aim to incorporate the \mirroreye~functionality into a physical robot to further examine the expressive potential that this approach may afford in coordination with physical action.

\subsection{Conclusion}
The introduction of a reflection-like overlay on the display of virtual eye models (\mirroreyes) significantly improved reference identification accuracy compared to a pure, non-reflective eye model without affecting identification speed.
In congruence with this accuracy gain, the subjective user experience in a game-like interaction with the eye models significantly improved with the \mirroreyes~compared to conditions that either lacked the reflection capability or the eye-like appearance. These results suggest that the benefits associated with \mirroreyes~cannot be attributed to the underlying eye model or the reflection alone, but only to their combination. Subsequent research should investigate the transfer of such effects to the reflections of other classes of gaze targets, explore options for a selective modulation of mirroring for attention guidance, and ascertain meaningful coordination with degrees of freedom available to more physically integrated machine agents.

\section*{Acknowledgments}
We thank Christian Arzate, Yotam Sechayk, and Heiko Wersing for their thoughtful discussions, which were instrumental in shaping the user study design.

\bibliographystyle{ieeetr}
\bibliography{mirror_eye_lit} 

\begin{thebibliography}{10}
\providecommand{\url}[1]{#1}
\csname url@samestyle\endcsname
\providecommand{\newblock}{\relax}
\providecommand{\bibinfo}[2]{#2}
\providecommand{\BIBentrySTDinterwordspacing}{\spaceskip=0pt\relax}
\providecommand{\BIBentryALTinterwordstretchfactor}{4}
\providecommand{\BIBentryALTinterwordspacing}{\spaceskip=\fontdimen2\font plus
\BIBentryALTinterwordstretchfactor\fontdimen3\font minus \fontdimen4\font\relax}
\providecommand{\BIBforeignlanguage}[2]{{%
\expandafter\ifx\csname l@#1\endcsname\relax
\typeout{** WARNING: IEEEtran.bst: No hyphenation pattern has been}%
\typeout{** loaded for the language `#1'. Using the pattern for}%
\typeout{** the default language instead.}%
\else
\language=\csname l@#1\endcsname
\fi
#2}}
\providecommand{\BIBdecl}{\relax}
\BIBdecl

\bibitem{kobayashi1997unique}
H.~Kobayashi and S.~Kohshima, ``Unique morphology of the human eye,'' \emph{Nature}, vol. 387, no. 6635, pp. 767--768, 1997.

\bibitem{Fang2024}
Y.~Fang, J.~M. P{\'e}rez-Moler{\'o}n, L.~Merino, S.-L. Yeh, S.~Nishina, and R.~Gomez, ``Enhancing social robot's direct gaze expression through vestibulo-ocular movements,'' \emph{Advanced Robotics}, vol.~38, no. 19-20, pp. 1457--1469, 2024.

\bibitem{Langton1999}
S.~R. Langton and V.~Bruce, ``Reflexive visual orienting in response to the social attention of others,'' \emph{Visual cognition}, pp. 541--567, 1999.

\bibitem{Langton2000}
S.~R. Langton, R.~J. Watt, and V.~Bruce, ``Do the eyes have it? cues to the direction of social attention,'' \emph{Trends in cognitive sciences}, vol.~4, no.~2, pp. 50--59, 2000.

\bibitem{Itier2007}
R.~J. Itier, C.~Villate, and J.~D. Ryan, ``Eyes always attract attention but gaze orienting is task-dependent: Evidence from eye movement monitoring,'' \emph{Neuropsychologia}, vol.~45, no.~5, pp. 1019--1028, 2007.

\bibitem{Driver1999}
J.~Driver~IV, G.~Davis, P.~Ricciardelli, P.~Kidd, E.~Maxwell, and S.~Baron-Cohen, ``Gaze perception triggers reflexive visuospatial orienting,'' \emph{Visual cognition}, vol.~6, no.~5, pp. 509--540, 1999.

\bibitem{Tomasello2007}
M.~Tomasello, B.~Hare, H.~Lehmann, and J.~Call, ``Reliance on head versus eyes in the gaze following of great apes and human infants: the cooperative eye hypothesis,'' \emph{Journal of human evolution}, vol.~52, no.~3, pp. 314--320, 2007.

\bibitem{Baron2001}
S.~Baron-Cohen, S.~Wheelwright, J.~Hill, Y.~Raste, and I.~Plumb, ``The “reading the mind in the eyes” test revised version: a study with normal adults, and adults with asperger syndrome or high-functioning autism,'' \emph{The Journal of Child Psychology and Psychiatry and Allied Disciplines}, vol.~42, no.~2, pp. 241--251, 2001.

\bibitem{Adolphs2005}
R.~Adolphs, F.~Gosselin, T.~W. Buchanan, D.~Tranel, P.~Schyns, and A.~R. Damasio, ``A mechanism for impaired fear recognition after amygdala damage,'' \emph{Nature}, vol. 433, no. 7021, pp. 68--72, 2005.

\bibitem{Hietanen2008}
J.~K. Hietanen, J.~M. Lepp{\"a}nen, M.~J. Peltola, K.~Linna-Aho, and H.~J. Ruuhiala, ``Seeing direct and averted gaze activates the approach--avoidance motivational brain systems,'' \emph{Neuropsychologia}, vol.~46, no.~9, pp. 2423--2430, 2008.

\bibitem{Senju2009}
A.~Senju and M.~H. Johnson, ``The eye contact effect: mechanisms and development,'' \emph{Trends in cognitive sciences}, vol.~13, no.~3, 2009.

\bibitem{Metta2008}
G.~Metta, G.~Sandini, D.~Vernon, L.~Natale, and F.~Nori, ``The icub humanoid robot: An open platform for research in embodied cognition,'' in \emph{Proceedings of the 8th Workshop on Performance Metrics for Intelligent Systems}, ser. PerMIS '08.\hskip 1em plus 0.5em minus 0.4em\relax New York, NY, USA: Association for Computing Machinery, 2008, p. 50–56.

\bibitem{Breazeal2008}
C.~Breazeal, M.~Siegel, M.~Berlin, J.~Gray, R.~Grupen, P.~Deegan, J.~Weber, K.~Narendran, and J.~McBean, ``Mobile, dexterous, social robots for mobile manipulation and human-robot interaction,'' in \emph{ACM SIGGRAPH 2008 new tech demos}.\hskip 1em plus 0.5em minus 0.4em\relax ACM, 2008, pp. 1--1.

\bibitem{Zaraki2014gazecontrol}
A.~Zaraki, D.~Mazzei, M.~Giuliani, and D.~De~Rossi, ``Designing and evaluating a social gaze-control system for a humanoid robot,'' \emph{IEEE Transactions on Human-Machine Systems}, vol.~44, no.~2, 2014.

\bibitem{admoni2017social}
H.~Admoni and B.~Scassellati, ``Social eye gaze in human-robot interaction: a review,'' \emph{Journal of Human-Robot Interaction}, 2017.

\bibitem{Gomez2018}
R.~Gomez, D.~Szapiro, K.~Galindo, and K.~Nakamura, ``Haru: Hardware design of an experimental tabletop robot assistant,'' in \emph{Proceedings of the 2018 ACM/IEEE International Conference on Human-Robot Interaction}, ser. HRI '18.\hskip 1em plus 0.5em minus 0.4em\relax New York, NY, USA: Association for Computing Machinery, 2018, p. 233–240.

\bibitem{Yoshida2022}
N.~Yoshida, S.~Yonemura, M.~Emoto, K.~Kawai, N.~Numaguchi, H.~Nakazato, S.~Otsubo, M.~Takada, and K.~Hayashi, ``Production of character animation in a home robot: A case study of lovot,'' \emph{International Journal of Social Robotics}, vol.~14, no.~1, pp. 39--54, 2022.

\bibitem{Kiilavuori2021roboteyecontact}
H.~Kiilavuori, V.~Sariola, M.~J. Peltola, and J.~K. Hietanen, ``Making eye contact with a robot: Psychophysiological responses to eye contact with a human and with a humanoid robot,'' \emph{Biological Psychology}, vol. 158, p. 107989, 2021.

\bibitem{Linnunsalo2023Intentionality}
S.~Linnunsalo, D.~Küster, S.~Yrttiaho, M.~J. Peltola, and J.~K. Hietanen, ``Psychophysiological responses to eye contact with a humanoid robot: Impact of perceived intentionality,'' \emph{Neuropsychologia}, vol. 189, p. 108668, 2023.

\bibitem{Krueger2017}
M.~Kr{\"u}ger, C.~B. Wiebel, and H.~Wersing, ``From tools towards cooperative assistants,'' in \emph{Proceedings of the 5th International Conference on Human Agent Interaction}, 2017, pp. 287--294.

\bibitem{Sendhoff2020}
B.~Sendhoff and H.~Wersing, ``Cooperative intelligence-a humane perspective,'' in \emph{2020 IEEE international conference on human-machine systems (ICHMS)}.\hskip 1em plus 0.5em minus 0.4em\relax IEEE, 2020, pp. 1--6.

\bibitem{Bhaskara2020transparency}
A.~Bhaskara, M.~Skinner, and S.~Loft, ``Agent transparency: A review of current theory and evidence,'' \emph{IEEE Transactions on Human-Machine Systems}, vol.~50, no.~3, pp. 215--224, 2020.

\bibitem{teyssier2021eyecam}
M.~Teyssier, M.~Koelle, P.~Strohmeier, B.~Fruchard, and J.~Steimle, ``Eyecam: Revealing relations between humans and sensing devices through an anthropomorphic webcam,'' in \emph{Proceedings of the 2021 CHI Conference on Human Factors in Computing Systems}, 2021, pp. 1--13.

\bibitem{Mori2012}
M.~Mori, K.~F. MacDorman, and N.~Kageki, ``The uncanny valley [from the field],'' \emph{IEEE Robotics and Automation Magazine}, vol.~19, no.~2, 2012.

\bibitem{fang2023designing}
Y.~Fang, L.~Merino, S.~Thill, and R.~Gomez, ``Designing visual and auditory attention-driven movements of a tabletop robot,'' in \emph{2023 32nd IEEE International Conference on Robot and Human Interactive Communication (RO-MAN)}, 2023, pp. 2232--2237.

\bibitem{Fang2024Roman}
Y.~Fang, J.~M. Pérez-Molerón, L.~Merino, and R.~Gomez, ``Enhancing human perception of direct gaze from a social robot through eye-head coordination,'' in \emph{2024 33rd IEEE International Conference on Robot and Human Interactive Communication (ROMAN)}, 2024.

\bibitem{Howard1995}
I.~P. Howard and B.~J. Rogers, \emph{Binocular vision and stereopsis}.\hskip 1em plus 0.5em minus 0.4em\relax Oxford University Press, USA, 1995.

\bibitem{Lee2008}
J.~C. Lee, ``Hacking the nintendo wii remote,'' \emph{IEEE Pervasive Computing}, vol.~7, no.~3, pp. 39--45, 2008.

\bibitem{bates2018head}
T.~Bates, J.~Kober, and M.~Gienger, ``Head-tracked off-axis perspective projection improves gaze readability of 3d virtual avatars,'' in \emph{SIGGRAPH Asia 2018}.\hskip 1em plus 0.5em minus 0.4em\relax Association for Computing Machinery (ACM), 2018, p.~29.

\bibitem{Delaunay2009}
F.~Delaunay, J.~De~Greeff, and T.~Belpaeme, ``Towards retro-projected robot faces: an alternative to mechatronic and android faces,'' in \emph{RO-MAN 2009-The 18th IEEE International Symposium on Robot and Human Interactive Communication}.\hskip 1em plus 0.5em minus 0.4em\relax Ieee, 2009, pp. 306--311.

\bibitem{python}
G.~Van~Rossum and F.~L. Drake~Jr, \emph{Python reference manual}.\hskip 1em plus 0.5em minus 0.4em\relax Centrum voor Wiskunde en Informatica Amsterdam, 1995.

\bibitem{opencv_library}
G.~Bradski, ``{The OpenCV Library},'' \emph{Dr. Dobb's Journal of Software Tools}, 2000.

\bibitem{mediapipe}
C.~Lugaresi, J.~Tang, H.~Nash, C.~McClanahan, E.~Uboweja, M.~Hays, F.~Zhang, C.-L. Chang, M.~Yong, J.~Lee, W.-T. Chang, W.~Hua, M.~Georg, and M.~Grundmann, ``Mediapipe: A framework for perceiving and processing reality,'' in \emph{Third Workshop on Computer Vision for AR/VR at IEEE Computer Vision and Pattern Recognition (CVPR)}, 2019.

\bibitem{schrepp2017design}
M.~Schrepp, J.~Thomaschewski, and A.~Hinderks, ``Design and evaluation of a short version of the user experience questionnaire (ueq-s),'' \emph{International Journal of Interactive Multimedia and Artificial Intelligence}, vol.~4, no.~6, pp. 103--108, 12/2017 2017.

\bibitem{brooke1996sus}
J.~Brooke \emph{et~al.}, ``Sus-a quick and dirty usability scale,'' \emph{Usability evaluation in industry}, vol. 189, no. 194, pp. 4--7, 1996.

\bibitem{Finstad2010umux}
K.~Finstad, ``The usability metric for user experience,'' \emph{Interacting with Computers}, vol.~22, no.~5, pp. 323--327, 05 2010.

\bibitem{hassenzahl2010needs}
M.~Hassenzahl, S.~Diefenbach, and A.~G{\"o}ritz, ``Needs, affect, and interactive products--facets of user experience,'' \emph{Interacting with computers}, vol.~22, no.~5, pp. 353--362, 2010.

\bibitem{pandas}
T.~pandas~development team, ``pandas-dev/pandas: Pandas,'' Feb. 2020.

\bibitem{matplotlib}
J.~D. Hunter, ``Matplotlib: A 2d graphics environment,'' \emph{Computing in Science \& Engineering}, vol.~9, no.~3, pp. 90--95, 2007.

\bibitem{seaborn}
M.~L. Waskom, ``seaborn: statistical data visualization,'' \emph{Journal of Open Source Software}, vol.~6, no.~60, p. 3021, 2021.

\bibitem{statsmodels}
S.~Seabold and J.~Perktold, ``statsmodels: Econometric and statistical modeling with python,'' in \emph{9th Python in Science Conference}, 2010.

\bibitem{scikit-posthocs}
M.~A. Terpilowski, ``scikit-posthocs: Pairwise multiple comparison tests in python,'' \emph{Journal of Open Source Software}, vol.~4, no.~36, 2019.

\bibitem{Vallat2018}
R.~Vallat, ``Pingouin: statistics in python,'' \emph{Journal of Open Source Software}, vol.~3, no.~31, p. 1026, Nov. 2018.

\bibitem{Mccarthy1997}
G.~McCarthy, A.~Puce, J.~C. Gore, and T.~Allison, ``Face-specific processing in the human fusiform gyrus,'' \emph{Journal of cognitive neuroscience}, vol.~9, no.~5, pp. 605--610, 1997.

\bibitem{besson2017}
G.~Besson, G.~Barragan-Jason, S.~J. Thorpe, M.~Fabre-Thorpe, S.~Puma, M.~Ceccaldi, and E.~J. Barbeau, ``From face processing to face recognition: Comparing three different processing levels,'' \emph{Cognition}, vol. 158, pp. 33--43, 2017.

\bibitem{George2001}
N.~George, J.~Driver, and R.~J. Dolan, ``Seen gaze-direction modulates fusiform activity and its coupling with other brain areas during face processing,'' \emph{Neuroimage}, vol.~13, no.~6, pp. 1102--1112, 2001.

\bibitem{Li2011}
Y.~H. Li and N.~Tottenham, ``Seeing yourself helps you see others.'' \emph{Emotion}, vol.~11, no.~5, p. 1235, 2011.

\bibitem{Mulkens2009}
S.~Mulkens and A.~Jansen, ``Mirror gazing increases attractiveness in satisfied, but not in dissatisfied women: A model for body dysmorphic disorder?'' \emph{Journal of Behavior Therapy and Experimental Psychiatry}, vol.~40, no.~2, pp. 211--218, 2009.

\bibitem{Barnier2019}
E.~M. Barnier and J.~Collison, ``Experimental induction of self-focused attention via mirror gazing: Effects on body image, appraisals, body-focused shame, and self-esteem,'' \emph{Body Image}, vol.~30, 2019.

\bibitem{Ypsilanti2020}
A.~Ypsilanti, A.~Robson, L.~Lazuras, P.~A. Powell, and P.~G. Overton, ``Self-disgust, loneliness and mental health outcomes in older adults: an eye-tracking study,'' \emph{Journal of affective disorders}, vol. 266, 2020.

\bibitem{Luo2019roboticeyespersonality}
L.~Luo, N.~Koyama, K.~Ogawa, and H.~Ishiguro, ``Robotic eyes that express personality,'' \emph{Advanced Robotics}, vol.~33, no. 7-8, 2019.

\bibitem{tao2023embodying}
Y.~Tao, C.~Y. Wang, A.~D. Wilson, E.~Ofek, M.~González-Franco, and A.~D. Wilson, ``Embodying physics-aware avatars in virtual reality,'' in \emph{CHI '23: Proceedings of the 2023 CHI Conference on Human Factors in Computing Systems}, April 2023.

\bibitem{Ruhland2015}
K.~Ruhland, C.~E. Peters, S.~Andrist, J.~B. Badler, N.~I. Badler, M.~Gleicher, B.~Mutlu, and R.~McDonnell, ``A review of eye gaze in virtual agents, social robotics and hci: Behaviour generation, user interaction and perception,'' \emph{Computer Graphics Forum}, vol.~34, no.~6, pp. 299--326, 2015.

\bibitem{yamashita2024personality}
J.~Yamashita, Y.~Takimoto, H.~Oishi, and T.~Kumada, ``How do personality traits modulate real-world gaze behavior? generated gaze data shows situation-dependent modulations,'' \emph{Frontiers in Psychology}, vol.~14, p. 1144048, 2024.

\bibitem{gomez2021developing}
R.~Gomez, Y.~Fang, S.~Thill, R.~Ragel, H.~Brock, K.~Nakamura, Y.~Vasylkiv, E.~Nichols, and L.~Merino, ``Developing a robot’s empathetic reactive response inspired by a bottom-up attention model,'' in \emph{International Conference on Social Robotics}.\hskip 1em plus 0.5em minus 0.4em\relax Springer, 2021.

\bibitem{oliveira2024differential}
M.~Oliveira, C.~Fernandes, F.~Barbosa, and F.~Ferreira-Santos, ``Differential correlates of fear and anxiety in salience perception: A behavioral and erp study with adolescents,'' \emph{Cognitive, Affective, \& Behavioral Neuroscience}, pp. 1--13, 2024.

\bibitem{Langton2008faces}
S.~R. Langton, A.~S. Law, A.~M. Burton, and S.~R. Schweinberger, ``Attention capture by faces,'' \emph{Cognition}, vol. 107, no.~1, 2008.

\bibitem{Gliga2009faces}
T.~Gliga, M.~Elsabbagh, A.~Andravizou, and M.~Johnson, ``Faces attract infants' attention in complex displays,'' \emph{Infancy}, vol.~14, no.~5, 2009.

\bibitem{Zheng2013robotmediatedjointattention}
Z.~Zheng, L.~Zhang, E.~Bekele, A.~Swanson, J.~A. Crittendon, Z.~Warren, and N.~Sarkar, ``Impact of robot-mediated interaction system on joint attention skills for children with autism,'' in \emph{2013 IEEE 13th International Conference on Rehabilitation Robotics (ICORR)}, 2013, pp. 1--8.

\bibitem{dawson2004early}
G.~Dawson, K.~Toth, R.~Abbott, J.~Osterling, J.~Munson, A.~Estes, and J.~Liaw, ``Early social attention impairments in autism: social orienting, joint attention, and attention to distress.'' \emph{Developmental psychology}, vol.~40, no.~2, p. 271, 2004.

\bibitem{charman2003jointattention}
T.~Charman, ``Why is joint attention a pivotal skill in autism?'' \emph{Philosophical Transactions of the Royal Society of London. Series B: Biological Sciences}, vol. 358, no. 1430, pp. 315--324, 2003.

\bibitem{Srinivasan2014headgazespeech}
V.~Srinivasan, C.~L. Bethel, and R.~R. Murphy, ``Evaluation of head gaze loosely synchronized with real-time synthetic speech for social robots,'' \emph{IEEE Transactions on Human-Machine Systems}, vol.~44, no.~6, pp. 767--778, 2014.

\end{thebibliography}

\end{document}